\documentclass{emulateapj} 
\usepackage{psfig} 

\received{}
\revised{}
\accepted{}

\shorttitle{Chandra and DEEP2 Galaxy Groups and Clusters} 

\shortauthors{Fang et al.}

\begin{document}

\title{Aegis: {\sl Chandra} Observation of DEEP2 Galaxy Groups and Clusters}

\author{Taotao~Fang\altaffilmark{1,2}, Brian F.~Gerke\altaffilmark{3},
David S.~Davis\altaffilmark{4,5}, Jeffrey A.~Newman\altaffilmark{6,7}, Marc
Davis\altaffilmark{1,3}, Kirpal Nandra\altaffilmark{8}, Elise
S.~Laird\altaffilmark{9}, David C.~Koo\altaffilmark{9}, Alison 
L.~Coil\altaffilmark{7,10}, Michael C.~Cooper\altaffilmark{1},
Darren J.~Croton\altaffilmark{1}, Renbin Yan\altaffilmark{1}}

\altaffiltext{1}{Department of Astronomy,
University of California, Berkeley, CA 94720}  

\altaffiltext{2}{Chandra fellow; fangt@astro.berkeley.edu}

\altaffiltext{3}{Department of Physics, University of California,
Berkeley, CA 94720} 

\altaffiltext{4}{Joint Center for Astrophysics, Department of Physics,
University of Maryland,
Baltimore County, 1000 Hilltop Circle, Baltimore, MD 21250,
ddavis@milkyway.gsfc.nasa.gov} 

\altaffiltext{5}{Laboratory for High Energy Astrophysics, Code 661,
  Greenbelt, MD 20771}  

\altaffiltext{6}{Institute for Nuclear and Particle
Astrophysics, Lawrence Berkeley National Laboratory, Berkeley, CA
94720}  

\altaffiltext{7}{Hubble fellow}

\altaffiltext{8}{Astrophysics Group, Imperial College London,
Blackett Lab., Prince Consort Road, London, SW7 2AW, UK}

\altaffiltext{9}{UCO/Lick Observatory, Department of Astronomy and
Astrophysics, University of California, Santa Cruz, CA 95064}

\altaffiltext{10}{Steward Observatory, University of
Arizona, Tucson, AZ 85721}

\begin{abstract}

We present a 200 $ksec$ {\sl Chandra} observation of seven
spectroscopically selected, high redshift ($0.75 < z < 1.03$) galaxy groups
and clusters discovered by the DEEP2 Galaxy Redshift Survey in the
Extended Groth Strip (EGS). X-ray emission at the locations of these systems
is consistent with background. The 3$\sigma$ upper limits on
the bolometric X-ray luminosities ($L_X$) of these systems put a
strong constraint on the relation between $L_X$ and the velocity
dispersion of member galaxies $\sigma_{gal}$ at $z\sim 1$; the DEEP2
systems have lower luminosity than would be predicted by the local
relation. Our result is consistent 
with recent findings that at high redshift, optically selected
clusters tend to be X-ray underluminous.  A comparison with mock
catalogs indicates that it is unlikely that this effect is entirely caused by
a measurement bias between $\sigma_{gal}$ and the dark matter velocity
dispersion. Physically, the DEEP2 systems may still be in the process of
forming and hence not fully virialized, or they may be deficient in
hot gas compared to local systems. We find only one possibly extended
source in this  {\sl Chandra} field, which happens to lie outside the
DEEP2 coverage. 

\end{abstract}


\section{Introduction}

Groups and clusters of galaxies are the most massive,
dynamically relaxed objects in the universe. Past studies have
focused on low to moderate redshifts (for 
reviews, see, e.g., \citealp{ros02,voi05}) due to the lack
of reliable 
samples at high redshift. Recently, progress has been made in
finding high-$z$ clusters and groups, largely due to
improved detection techniques and greater instrument sensitivities
over large fields. For instance, the method of observing red sequence
galaxies has proven to be efficient in 
identifying clusters with optical/near-IR imaging data (e.g., \citealp{gye00}). Also, {\sl Chandra} and {\sl XMM}-Newton have
started to reveal X-ray clusters at redshift around unity (e.g.,\citealp{ros02,voi05}). However, until now, no optically selected,
spectroscopic sample has existed with sufficient size, sampling
density, and redshift accuracy to identify
large numbers of groups and clusters at $z > 0.5$. 

\begin{figure*}[thb]
\centerline{\includegraphics[angle=0,scale=0.8]{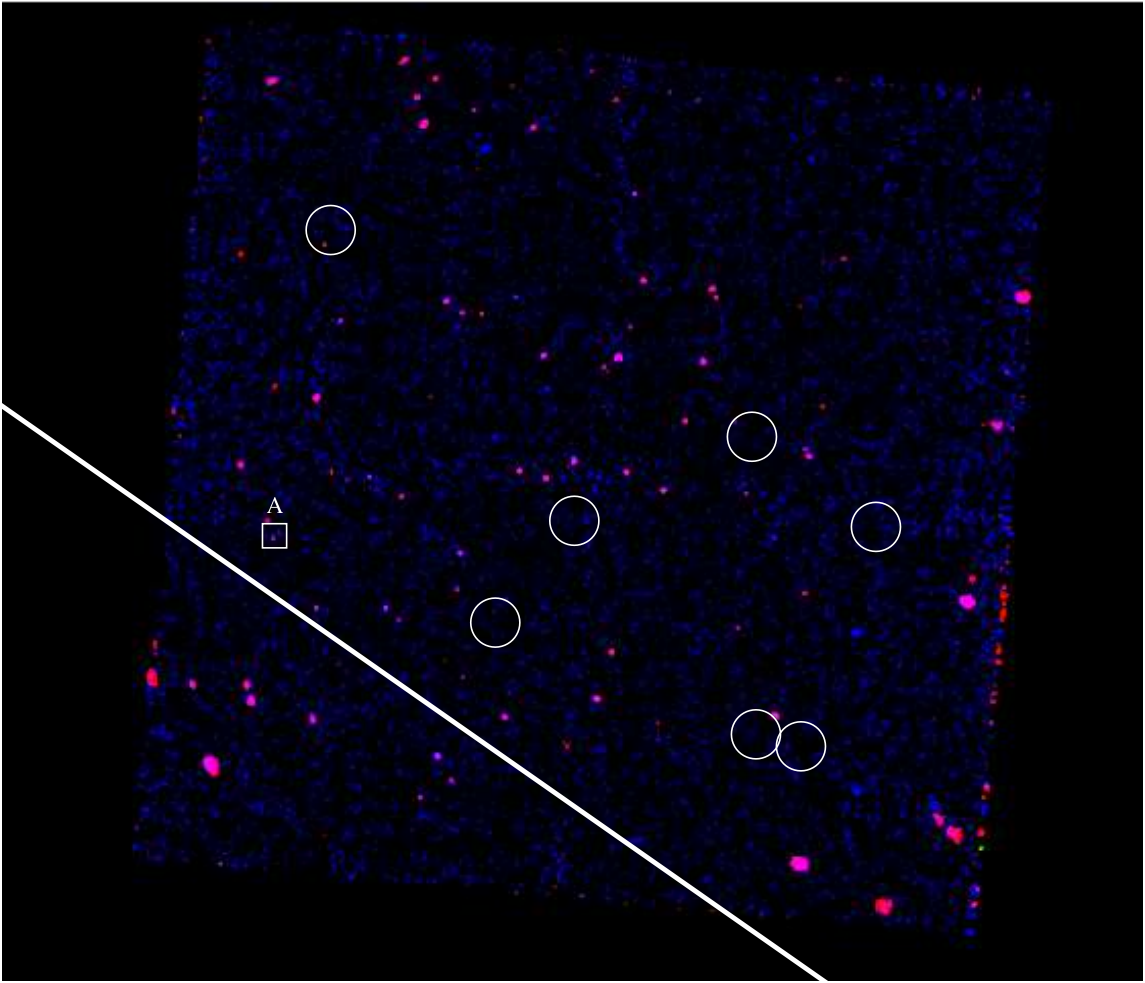}}
\caption[h]{True-color image of the {\sl Chandra} ACIS-I field,
with north to the top and east to the left. The
  image is color-coded so that photons with energies at 0.54 -- 2.05
  keV, 2.05 -- 3.50 keV, and 3.50 -- 7.00 keV are in red, green, and
  blue, respectively. The DEEP2 survey covers the region to
the north of the solid line. Circles represent DEEP2-identified
systems. The white box labeled ``A'' indicates a possibly extended source.} 
\label{f1}
\end{figure*}

The DEEP2 Galaxy Redshift Survey \citep{dav03} has provided the first large spectroscopic galaxy catalog focused on
$z\sim1$, based on observations of $\sim 5 \times 10^4$
galaxies. One of its four survey fields incorporates the Extended Groth Strip
(EGS), which is the target of a multiwavelength survey called
``All-wavelength Extended Groth Strip International Survey''
(AEGIS; \citealp{dav06}). DEEP2 has produced the largest well-defined,
spectroscopically selected sample of clusters and groups at high $z$ yet
\citep{ger05}.

In this {\it  Letter}, we investigate the X-ray properties of the
DEEP2 groups and clusters which overlap an archival 200 $ksec$ {\sl
Chandra} observation. The primary goal is to
understand the relation between optical and X-ray properties of
these systems at $z>0.7$. At low $z$, simple scaling relations between
X-ray quantities such as bolometric luminosity ($L_X$) and temperature
($T_X$) and optical quantities such as galaxy velocity dispersion
($\sigma_{gal}$) have been predicted and well studied (see, e.g., \citealp{whi97,all98,mar98,fin01,ett02}). However, high-$z$ observations reveal a
different trend: optically selected galaxy
clusters tend to be less X-ray luminous than those with similar
velocity dispersions at low redshifts (\citealp{cas94,bow97,hol97,lub02,lub04}), while the X-ray  
selected samples tend to follow the local $L_X$--$\sigma_{gal}$ relation
(see, e.g., \citealp{ebe01,val04}). The exact reason for the break-down of the
$L_X$--$\sigma_{gal}$ relation is still unclear. While previous
studies of optical-X-ray relations have been limited 
only to a few large clusters, DEEP2 allows us to identify both
clusters and groups at $z\sim 1$.

\section{{\sl Chandra} Data Analysis}

The AO3 {\sl Chandra} observation of the Groth-Westphal Strip (GWS), a
subset of EGS,  
was taken on August 2002 (see \citealp{nan05} for observation
details). This observation was taken using the imaging mode of the Advanced 
CCD Imaging Spectrometer (ACIS-I), which utilizes four
CCD chips with a total field of view of $\sim
17\arcmin\times17\arcmin$.  A total of $\sim 200\ ksec$
exposure time was obtained over three separate observations, with
IDs of \#3305 (30 $ksec$), \#4357 (85 $ksec$) and \#4365
(85 $ksec$). Each observation was processed with the Chandra
Interactive Analysis of Observations (CIAO) software package version
3.2\footnote{See http://asc.harvard.edu/ciao/.}. Specifically, we
first identified and rejected hot pixel and afterglow events in the level
1 events file to create a new level 2 events file, using the CIAO tool
``acis\_run\_hotpix''. We also removed bad pixels using
observation-specific bad pixel files, selecting events with standard
grades of 0, 2, 3, 4, 6, only, and cleaned the dataset
of periods of anomalous background levels caused by strong
background flares. The final level 2 events file was created by
co-adding the three observations. Figure~\ref{f1} shows a
true-color image of the {\sl Chandra} ACIS-I field after correction 
by exposure maps. The image is color-coded so that photons with
energies at 0.54 -- 2.05 keV, 2.05 -- 3.50 keV, and 3.50 -- 7.00 keV
are shown as red, green, and blue, respectively. 

We applied the Voronoi Tessellation and Percolation (VTP) algorithm
\citep{ebe93} to detection sources, using
the {\sl Chandra} tool ``vtpdetect''. While the VTP algorithm can be slow
for large numbers of photons, it does have the advantage of finding
faint, low surface-brightness features.  We required that all detected
sources should have at least 40 net photons, and set the maximum
probability of being a false source to be 
$10^{-6}$. We searched for potentially extended sources in three
bands using VTP. This search algorithm yielded a total of 65 sources in
the full band (0.5 -- 7 keV), 45 in the soft band (0.5 -- 2 keV), and
29 in the hard band (2 -- 7 keV). The region corresponding to each
detected source is represented by an ellipse with a semi-major axis of
$a$ and a semi-minor axis $b$.   

To examine whether these sources are truly extended, we compared the
size of the output ellipses with the size of instrument point spread
function (PSF) at their corresponding off-axis angles. We computed the
PSF using the MARX simulator \footnote{See
  http://space.mit.edu/ASC/MARX/.}. Specifically, single-energy  
photons at 1.5 keV were injected at different off-axis angles from the
aim-point of ACIS-I, and then encircled energies at various
percentage levels are calculated. We find that most sources have
sizes from VPF smaller than the 95\% encircled radius, indicating that they 
are indeed point sources. 

The only exception is source A at (RA, DEC)=(14:18:21.832, +52:26:5)
(J2000), which is nearly twice as large as its predicted  95\% radius
and is designated by the white box 
in Figure~\ref{f1}. Visual inspection of this source in the exposure-corrected
map shows that it indeed appears extended, although we cannot exclude the
possibility that it is a mix of 2 -- 3 faint point sources. The
significance of the detection is $\sim 6\sigma$. Assuming the emission
is from a weak extended source, we extracted the spectrum and fit it
using the "APEC" thermal emission model with the software package
XSPEC v12.0 \footnote{See
  http://heasarc.gsfc.nasa.gov/docs/xanadu/xspec/.}. We fixed the
neutral hydrogen absorption at the Galactic level, and assumed an
abundance  of 0.3 solar -- the abundance that is typically found in
the local galaxy groups and clusters. Due
to the very limited number of net counts ($\sim$ 75 photons in the
encircled region), we cannot constrain the temperature. Fixing it at 1
keV, we obtain a soft band absorbed flux (0.5 -- 2 keV) of $\sim 1.54 \times
10^{-15}\rm\ ergs\ cm^{-2}\ s^{-1}$.  

\section{DEEP2 Group Catalog}

Details of the DEEP2 group sample and group finding methods can be
found in \citet{ger05}. Here we 
briefly summarize information that is relevant to this {\it
Letter}. Groups were identified using the Voronoi-Delaunay method
(VDM, \citealp{mar02}). To insure a uniform group sample, we
trimmed the DEEP2 sample in EGS to match the target selection rate and
color cut used to select galaxies at $z>0.7$ in the remaining DEEP2
fields.  The group finding algorithm has been optimized using the mock
catalogs presented in \citet{yan04}.

We measure galaxy velocity dispersions in these groups using the
``gapper'' estimator (see, e.g., \citealp{bee90}).  For the sample
used in this paper, we require a minimum velocity dispersion of 300
$\rm km\ s^{-1}$, above which the group velocity function is recovered
well in mock catalogs \citep{ger05}. This velocity dispersion cutoff roughly corresponds to a
virialized halo mass of $\sim 10^{13}\ M_{\odot}$ at $z\sim 1$ with
our adopted cosmology. \footnote{We adopt a standard $\Lambda$CDM
  cosmological model with a matter density of 
$\Omega_M = 0.3$ and a cosmological constant of $\Omega_\Lambda =
0.7$, and a Hubble constant of $H_0 = 100 h\ \rm km\
s^{-1}Mpc^{-1}$, where $h=0.7$.} We also require systems to have more than two
member galaxies ($N_g > 2$). While in general errors on the measured
velocity dispersion could be quite large for systems with small
$N_g$ (see discussion in \S~5), we
find that such errors are too large to have any
constraining power at all when $N_g=2$. With these cuts, we find a 
total of 7 groups and clusters in the DEEP2 data overlapping the
{\sl Chandra} pointing, with redshifts $0.75 < z < 1.03$. In
Figure~\ref{f1} we show the positions of the DEEP2 
systems as white circles, with center indicating the mean position of
member galaxies and radius 30\arcsec. This 
radius roughly corresponds to a physical distance of $\sim 250$
kpc, a typical core radius for galaxy groups and clusters at high
redshift. 

\section{Results}

\subsection{High-$z$ $L_X - \sigma_{gal}$ relation}

None of the DEEP2 groups was detected directly by VPF.  However, we
can place upper limits on their X-ray luminosity. Using  Poisson
statistics, we estimate the 3$\sigma$ fluctuation in the X-ray
background within a 30\arcsec radius aperture and take this as a
3$\sigma$ upper limit on the X-ray photon counts from each
system. Assuming thermal Bremsstrahlung radiation, we can then convert
photon counts to flux using PIMMS
\footnote{Portable, Interactive Multi-Mission Simulator, see
  http://cxc.harvard.edu/toolkit/pimms.jsp}. To obtain flux, we also
  need a temperature estimate. Here we assume these systems follow the
  local,  well-calibrated $\sigma-T_X$ relationship (see, e.g., Xue \&
  Wu~2000). While as \citet{lub04} reported, the high-z clusters are
  generally cooler, we found that for fixed photon counts, X-ray flux
  is not a very sensitive function of temperature: by lowering the
  temperature by a factor of 5, the flux varies by less than
  20\%. Knowing flux and redshift, we can then obtain the intrinsic
  X-ray luminosity of each galaxy group.

\begin{figure}[t]
\centerline{\includegraphics[angle=90,scale=0.4]{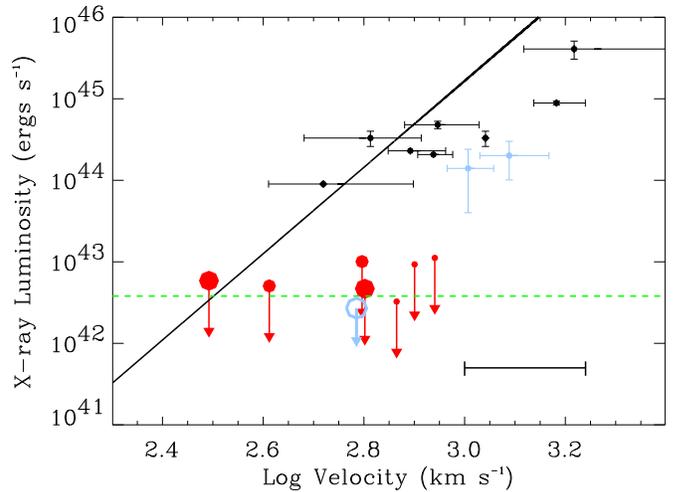}}
\caption[h]{Bolometric X-ray luminosity vs. velocity
  dispersion. Filled red circles are 3$\sigma$ upper limits on $L_X$
  for the 
  DEEP2 systems, with their mean $1\sigma$ $\sigma_{gal}$
uncertainty shown by the error bar at the bottom right. The size of
  each circle is
  proportional to the system richness, ranging from $N_g=3$ to 6.
The horizontal dashed green line indicates the $3\sigma$ upper
  limit from the stacked data. The solid dark line shows the local
relation from \citet{xue00}. High-$z$ ($z > 0.7$) X-ray selected
  clusters (black 
circles with error bars) are taken from \citealp{don99,gio99,ebe01,sta01,val04}. Optically selected
  clusters are taken from \citet{lub02} (blue circles with
  error bars) and \citet{bau02} (open blue circle with arrow).}
\label{f2}
\end{figure}

In Figure~\ref{f2} we plot the bolometric X-ray luminosity as a
function of velocity dispersion. The filled red circles are the 3$\sigma$
upper limits on $L_X$ for the DEEP2 systems. The size of the circles is
proportional to $N_g$, i.e., the smallest circle corresponds to 3
members, and the largest corresponds to 6 members. We compute
$\sigma_{gal}$ errors by using Monte-Carlo realizations to obtain
the distribution of true dispersions, given a richness and a
dispersion measured with the ``gapper'' estimator; the mean
uncertainty for the systems considered here is shown by the
horizontal error bar at the bottom right of Figure~\ref{f2}. 

The dark line in Figure~\ref{f2} is the local $L_X$
-- $\sigma$ relation from \citet{xue00}. Since DEEP2 systems and
cluster sample in \citet{xue00} have similar $\sigma_{gal}$, we use their
relation $L_X \propto \sigma^{5.30}$ that is appropriate for clusters of
galaxies. Clearly, the
DEEP2 systems are X-ray underluminous compared to the local
$L_X-\sigma_{gal}$ relation, and this trend is more apparent for
systems with larger velocity dispersions. This
result is robust to changes
in the adopted system radius. By
increasing the radius to $30\arcsec$ to $40\arcsec$, the 3$\sigma$
upper limit in $L_X$ increases by $\sim 35$\% only. We also have
stacked {\sl Chandra} images at their corresponding DEEP2 system
positions to see whether we can 
find enhanced X-ray emission. The stacked image is consistent  
with background emission and provides a stronger constraint on the
upper limit of $L_X$ (shown by the dashed green line in
Figure~\ref{f2}).

We overplot results from previous observations in
Figure~\ref{f2}. X-ray selected clusters are plotted in black and
optically selected clusters are plotted in blue. The open blue circle
with an upper 
limit is taken from the $\sim 1$ Ms observation in the {\sl Chandra}
Deep Field north \citep{bau02}, in which an optically identified
cluster at $z \approx 0.85$ was not detected in X-ray. It is quite
clear that at high-$z$, 
while optically selected clusters are X-ray underluminous, X-ray
selected systems more or less follow the local
$L_X - \sigma_{gal}$ relation. 

\subsection{Bias between galaxy and dark matter velocity dispersions}

How well does the measured $\sigma_{gal}$ reflect the dynamic state of
high-$z$ galaxy clusters and groups? A recent weak-lensing
measurement of the optically selected cluster Cl~1604+4304 at
$z=0.9$ shows that galaxies are indeed good tracers of the dark
matter, at least in this system 
\citep{mar05}. However, $\sigma_{gal}$ measurements will tend to be
higher than the the velocity dispersion of the dark matter
($\sigma_{DM}$) due to an Eddington-like bias: low-$\sigma_{DM}$
groups are much more common than high-$\sigma_{DM}$ ones, so given the
large velocity dispersion errors when $N_g$ is small, it is more
likely that a system above our velocity dispersion cutoff limit has a
$\sigma_{gal}$ which was scattered up from its true $\sigma_{DM}$
value than one which was scattered down.  

\begin{figure}[t]
\centerline{\includegraphics[angle=0,scale=0.4]{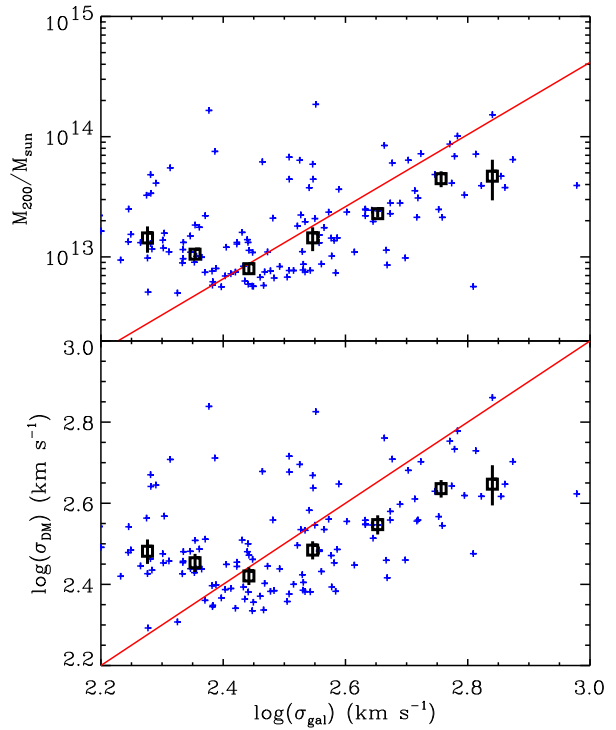}}
\caption[h]{Halo mass ($M_{200}$, top panel) and dark matter velocity
dispersion ($\sigma_{DM}$, bottom panel) as a function of galaxy
velocity
dispersion ($\sigma_{gal}$). The red lines in the top and bottom
panels represent the theoretical prediction of $M_{200}$
vs. $\sigma_{gal}$ \citep{nda02}, and $\sigma_{DM} = \sigma_{gal}$,
respectively. Blue crosses are systems identified in the mock catalogs
of \citet{yan04},
and the open squares with error bars 
indicate the medians and standard error of $M_{200}$ or
$\sigma_{DM}$ for equal-size bins in $\log(\sigma_{gal})$, with a
bin size of 0.1.}
\label{f3}
\end{figure}

To test this, we have looked at systems in the DEEP2 mock catalogs \citep{yan04} with $2 < N_g < 7$. In Figure~\ref{f3}, we plot the total mass within an overdensity of $\sim
200$ ($M_{200}$, top panel) and dark matter velocity dispersion
($\sigma_{DM}$, bottom panel) as a function of
$\sigma_{gal}$. Clearly, a bias exists at high $\sigma_{gal}$ end: for
the same system $\sigma_{gal}$ tends to be higher than $\sigma_{DM}$,
and the system mass inferred from $\sigma_{gal}$ is
overestimated. Such bias typically is not seen at low redshift. At low
$z$, simulations show that $\sigma_{gal}$ in general follows
$\sigma_{DM}$ quite well, although some bias does exist at $\sim$10\% level for
systems with less than 10 member galaxies (see, e.g., \citet{dav02})

While this bias can partly explain why the DEEP2 systems are X-ray
faint, it is unlikely that the non-detection of X-ray emission from
all seven systems is caused by this effect. Based on the 3$\sigma$
luminosity limit calculated in $\S 2$, non-detection would require
that all seven systems have
$\sigma_{DM} \lesssim 350\rm\ km\ s^{-1}$ if the local $L_X -
\sigma_{gal}$ relation applies. We have performed a Monte-Carlo
test by taking random sets of seven systems with measured velocity
dispersions $300 < \sigma_{gal} < 1000\rm\ km\ s^{-1}$ in the mock
catalog, and determine the chance that none of those systems have
$\sigma_{DM} > 350\rm\ km\ s^{-1}$ is only 0.14\%. So with
$\sim 3\sigma$ confidence the DEEP2 systems must be X-ray
underluminous.

\section{Discussion}

In this ${\it Letter}$,  we find evidence 
that spectroscopically selected galaxy groups and clusters at high
 redshift are X-ray underluminous; i.e., the upper limits of their
 X-ray luminosities are significantly below predictions from the local
 $L_X-\sigma_{gal}$ relation, unlike X-ray selected clusters at these
 redshifts. This confirms previous results showing that 
 optically selected clusters at high $z$ are weak X-ray sources.  

While bias in the $\sigma_{gal}$ measurement may contribute partly
to the non-detection, it is unlikely to be the sole explanation.
Physically, there are two basic reasons 
why spectroscopically selected systems at $z\sim1$ 
might fall below the $L_X-\sigma_{gal}$ relation followed by X-ray
selected clusters at these redshifts.  The first is
that the DEEP2 systems may be dynamically younger 
objects than those clusters which are easily detected in the X-ray.  If
the DEEP2 systems are are not yet fully virialized, their gas may tend
to be in cooler subclumps rather than at the temperature predicted from
spherical-collapse models, and their galaxy velocity distribution may
be multimodal rather than Gaussian, increasing the measured
$\sigma_{gal}$ (see, e.g., \citealp{fre96,bow97,lub04}).

An alternative is that these galaxy groups and clusters may be
deficient in the hot gas that is necessary to produce detectable X-ray 
emission.  DEEP2 systems do appear to be generally above the
threshold mass beyond which cooling is predicted to be inefficient
and a hot atmosphere forms (\citealp{bir03,ker05,cro06}). However, they may have passed that threshold only
relatively recently in their history and may still be in the process
of building a hot halo of substance. In this case $L_X$ would be
expected to remain low until a sufficient amount of hot gas had been
accreted. We note that the average blue galaxy fraction is higher in
DEEP2 groups than in local systems, reflecting the more common
presence of the cold gas which feeds star formation \citep{ger06}.
al. 2006, in prep.)  

Our {\sl Chandra} observation is $\sim 200\ ksec$. Two deeper
($\sim 1 Msec$) exposures, the {\sl Chandra} Deep Field South and
North, yield a total of 18 \citep{gia02} and 6 \citep{bau02} extended
sources, respectively. Most of the sources are early type galaxies,
poor groups and clusters. However, an early analysis on the {\sl
  Chandra} Deep Field North with similar exposure did not detect any
extended emission \citep{hor01}.

Our study highlights the importance of understanding groups and
clusters of galaxies at high redshift. These systems present very
different properties when compared with their counterparts at low
redshift, indicating that complex processes govern the formation
and evolution of large scale structures, which in turn have
significant impact on galaxy evolution.  We plan to investigate
larger DEEP2 samples in the future to advance this work.  

{\it Acknowledgements:} We thank Martin White, Joanne Cohn and Sandy
Faber for
useful discussions and comments. TF was supported by the NASA through {\sl
Chandra} Postdoctoral Fellowship Award Number PF3-40030 issued by the
{\sl Chandra} X-ray Observatory Center, which is operated by the
Smithsonian Astrophysical Observatory for and on behalf of the NASA
under contract NAS~8-39073. Also see Davis et al.~(2006) in this volume
for full acknowledgements.


\begin{thebibliography}{}

\bibitem[Allen \& Fabian(1998)]{all98} Allen, S.~W.,
\& Fabian, A.~C.\ 1998, \mnras, 297, L57

\bibitem[Bauer et al.(2002)]{bau02} Bauer, F.~E., et al.\ 
2002, \aj, 123, 1163

\bibitem[Beers et al.(1990)]{bee90} Beers, T.~C., Flynn, K., 
\& Gebhardt, K.\ 1990, \aj, 100, 32 

\bibitem[Birnboim \& Dekel(2003)]{bir03} Birnboim, Y., \& Dekel,
A.\ 2003, \mnras, 345, 349

\bibitem[Bower et al.(1997)]{bow97} Bower, R.~G., Castander, 
F.~J., Ellis, R.~S., Couch, W.~J., \& Boehringer, H.\ 1997, \mnras, 291, 
353 

\bibitem[Castander et al.(1994)]{cas94} Castander, F.~J., 
Ellis, R.~S., Frenk, C.~S., Dressler, A., \& Gunn, J.~E.\ 1994,
\apjl, 424, L79

\bibitem[Croton et al.(2006)]{cro06} Croton, D.~J., et al.\ 2006,
\mnras, 365, 11

\bibitem[Dav{\'e} et al.(2002)]{dav02} Dav{\'e}, R., Katz, 
N., \& Weinberg, D.~H.\ 2002, \apj, 579, 23 

\bibitem[Davis et al.(2003)]{dav03} Davis, M., et al.\ 2003, 
\procspie, 4834, 161 

\bibitem[Davis et al.(2006)]{dav06} Davis, M., et al.\ 2006, ApJL, submitted (this volume)

\bibitem[Donahue et al.(1999)]{don99} Donahue, M., Voit, 
G.~M., Scharf, C.~A., Gioia, I.~M., Mullis, C.~R., Hughes, J.~P., \& 
Stocke, J.~T.\ 1999, \apj, 527, 525

\bibitem[Ebeling \& Wiedenmann(1993)]{ebe93} Ebeling, H., \&
Wiedenmann, G.\ 1993, \pre, 47, 704

\bibitem[Ebeling et al.(2001)]{ebe01} Ebeling, H., Jones, 
L.~R., Fairley, B.~W., Perlman, E., Scharf, C., \& Horner, D.\ 2001, \apjl, 
548, L23

\bibitem[Ettori et al.(2002)]{ett02} Ettori,
S., De Grandi, S., \& Molendi, S.\ 2002, \aap, 391, 841

\bibitem[Finoguenov et al.(2001)]{fin01} Finoguenov, A., 
Reiprich, T.~H., Bohringer, H.\ 2001, \aap, 368, 749 

\bibitem[Frenk et al.(1996)]{fre96} Frenk, C.~S., Evrard, 
A.~E., White, S.~D.~M., \& Summers, F.~J.\ 1996, \apj, 472, 460 

\bibitem[Gerke et al.(2005)]{ger05} Gerke, B.~F., et al.\  2005,
\apj, 625, 6

\bibitem[Gerke et al.(2006)]{ger06} Gerke, B.~F., et al.\ 2006,
\apj, in preparation

\bibitem[Giacconi et al.(2002)]{gia02} Giacconi, R., et al.\ 
2002, \apjs, 139, 369 

\bibitem[Gioia et al.(1999)]{gio99} Gioia, I.~M., Henry, 
J.~P., Mullis, C.~R., Ebeling, H., \& Wolter, A.\ 1999, \aj, 117, 2608

\bibitem[Gladders \& Yee(2000)]{gye00} Gladders, M.~D., \& 
Yee, H.~K.~C.\ 2000, \aj, 120, 2148 

\bibitem[Holden et al.(1997)]{hol97} Holden, B.~P., Romer, 
A.~K., Nichol, R.~C., \& Ulmer, M.~P.\ 1997, \aj, 114, 1701

\bibitem[Hornschemeier et al.(2001)]{hor01} Hornschemeier, 
A.~E., et al.\ 2001, \apj, 554, 742

\bibitem[Kere{\v s} et al.(2005)]{ker05} Kere{\v s}, D., Katz, N.,
Weinberg, D.~H., \& Dav{\'e}, R.\ 2005, \mnras, 363, 2

\bibitem[Lubin et al.(2004)]{lub04} Lubin, L.~M., Mulchaey, 
J.~S., \& Postman, M.\ 2004, \apjl, 601, L9 
 
 \bibitem[Lubin et al.(2002)]{lub02} Lubin, L.~M., Oke, J.~B., 
\& Postman, M.\ 2002, \aj, 124, 1905 

\bibitem[Margoniner et al.(2005)]{mar05} Margoniner, V.~E., 
Lubin, L.~M., Wittman, D.~M., \& Squires, G.~K.\ 2005, \aj, 129, 20

\bibitem[Markevitch(1998)]{mar98} Markevitch, M.\ 1998, \apj, 504, 27 

\bibitem[Marinoni et al.(2002)]{mar02} Marinoni, C., Davis, 
M., Newman, J.~A., \& Coil, A.~L.\ 2002, \apj, 580, 122 

\bibitem[Nandra et al.(2005)]{nan05} Nandra, K., et al.\ 
2005, \mnras, 356, 568

\bibitem[Newman \& Davis(2002)]{nda02} Newman, J.~A., \& 
Davis, M.\ 2002, \apj, 564, 567 

\bibitem[Rosati et al.(2002)]{ros02} Rosati, P., Borgani, S., 
\& Norman, C.\ 2002, \araa, 40, 539 

\bibitem[Stanford et al.(2001)]{sta01} Stanford, S.~A., 
Holden, B., Rosati, P., Tozzi, P., Borgani, S., Eisenhardt, P.~R., \& 
Spinrad, H.\ 2001, \apj, 552, 504 

\bibitem[Valtchanov et al.(2004)]{val04} Valtchanov, I., et 
al.\ 2004, \aap, 423, 75 

\bibitem[Voit(2005)]{voi05} Voit, G.~M\ 2005, Rev. Mod. Phys., 77, 207

\bibitem[White et al.(1997)]{whi97} White, D.~A.,
Jones, C., \& Forman, W.\ 1997, \mnras, 292, 419 
 
\bibitem[Xue \& Wu(2000)]{xue00} Xue, Y.-J., \& Wu, X.-P.\  2000, \apj, 538, 65 

\bibitem[Yan et al.(2004)]{yan04} Yan, R., White, M., \& Coil,
A.~L.\ 2004, \apj, 607, 739  

\end{thebibliography}
\end{document}